\documentclass[aps,twocolumn,showpacs,pra,superscriptaddress,floatfix,longbibliography]{revtex4-1}

\usepackage{amsmath,amsthm,amsfonts,amssymb,times,bbm,graphicx,color,epsfig,bm}
\newtheorem{thm}{Theorem}
\newcommand{\bra}[1]{\mbox{$\langle #1 |$}}
\newcommand{\ket}[1]{\mbox{$| #1 \rangle$}}

\newcommand{\bd}[1]{\boldsymbol{#1}}

\newcommand{\R}{{\mathbb{R}}}

\begin{document}

\title{Universal upper bounds on the Bose-Einstein condensate and the Hubbard star}

\author{Felix Tennie}
%\email{felix.tennie@physics.ox.ac.uk}
\affiliation{Clarendon Laboratory, University of Oxford, Parks Road, Oxford OX1 3PU, United Kingdom}

\author{Vlatko Vedral}
\affiliation{Clarendon Laboratory, University of Oxford, Parks Road, Oxford OX1 3PU, United Kingdom}
\affiliation{Centre for Quantum Technologies, National University of Singapore, 3 Science Drive 2, Singapore 117543}

\author{Christian Schilling}
\email{christian.schilling@physics.ox.ac.uk}
\affiliation{Clarendon Laboratory, University of Oxford, Parks Road, Oxford OX1 3PU, United Kingdom}

\date{\today}

\begin{abstract}
For $N$ hard-core bosons on  an arbitrary lattice with $d$ sites and independent of additional interaction terms we prove that the
hard-core constraint itself already enforces a universal upper bound on the Bose-Einstein condensate given by $N_{max}=(N/d)(d-N+1)$. This bound can only be attained for one-particle states $\ket{\varphi}$ with equal amplitudes with respect to the hard-core basis (sites) and when the corresponding $N$-particle state $\ket{\Psi}$ is maximally delocalized. This result is generalized to the maximum condensate possible within a given sublattice. We observe that such maximal local condensation is only possible if the mode entanglement between the sublattice and its complement is minimal.
We also show that the maximizing state $\ket{\Psi}$ is related to the ground state of a bosonic `Hubbard star' showing Bose-Einstein condensation.
\end{abstract}


\maketitle
\section{Introduction}\label{sec:intro}

Bose-Einstein condensation (BEC) is one of the most fascinating quantum phenomena.
It was predicted almost one century ago following from the work by Bose \cite{Bose1924} and Einstein \cite{Einstein1924, Einstein1925} on the quantum gas of noninteracting  bosons. A lot of effort has been devoted ever since to investigate and understand the role of
particle-particle interactions on the occurrence of BEC. In particular, since the concept of one-particle energy states does not make sense anymore a more general criterion for BEC was provided by Onsager and Penrose \cite{Penrose1956} for the case of interacting bosons: A system of $N$ bosons exhibits BEC whenever its largest eigenvalue of the corresponding one-particle reduced density matrix $\rho(\vec{x},\vec{x}')$ is proportional to $N$.
Such a macroscopic occupancy is closely related to long-range order of the `off-diagonal' elements of  $\rho(\vec{x},\vec{x}')$ \cite{Yang1962}. Application of those two criteria to homogeneous gases has shown that BEC can also exist in the presence of interactions in three and more spatial dimensions (see, e.g., the reviews \cite{Shi1998,Andersen2004}).
The experimental discovery of BEC for trapped ultra-cold gases \cite{Anderson1995,Davis1995} has strongly revived the study of BEC for both, translationally invariant and trapped systems \cite{Dalfovo1999}. In this context hard-core bosons (HCB), originally introduced as a lattice model for liquid Helium II to investigate superfluidity \cite{Matsubara1956,Matsuda1957}, gained tremendous relevance: They can be realized experimentally, as demonstrated the first time in Ref.~\cite{Paredes2004}, by tuning the interaction between ultracold atoms at the Feshbach resonance to a repulsive contact potential \cite{Bloch2008, Chin2010, Weidemueller2011, Zuern2012}.

An interesting observation was made by Girardeau \cite{Girardeau1960} for \emph{one-dimensional} systems. The energy spectrum and other phase-independent quantities like density correlation functions, always coincide for spinless HCB and the analogous system of spinless fermions.
Yet, since the one-particle reduced density matrix $\rho(x,x')$ is phase-sensitive, the question whether occupation numbers can exceed the value one or may even describe BEC is \emph{a priori} non-trivial for HCB in contrast to fermions. In Refs.~\cite{Schultz1963, Lenard1964} the largest occupation number for $N$ HCB (without further interactions besides the impenetrability) in one dimension was shown to be proportional to $\sqrt{N}$ implying the absence of BEC. The same results hold for the case of HCB in an external harmonic trap \cite{Forrester2003a,Forrester2003b} and for the corresponding lattice analogs \cite{RigolHCB1d2005a}.

These \emph{specific} results on the absence of BEC for hard-core bosons even at zero temperature motivate a couple of questions:
Is the hard-core constraint itself already so restrictive that no (or no complete) BEC is possible \emph{independent} of the external potential and the type of particle-particle interaction? In particular for the case of lattice HCB, what is the maximal possible occupation number $N_{max}$ as function of the particle number $N$ and the number $d$ of available sites? How do the one-particle quantum states $\ket{\varphi_{max}}$ allowing for such a maximal occupation number look like and what is the form of the corresponding $N$-HCB state $\ket{\Psi_{max}}$ attaining this occupancy $N_{max}$ of $\ket{\varphi_{max}}$? In this paper we are going to answer all those questions. In addition, in Sec.~\ref{sec:Hubbard}, we will propose a physical model for HCB which allows the realization of a state with a macroscopically large occupation number saturating our universal upper bound. Let us first introduce some elementary concepts relevant for our work.

\section{Hard-core bosons: Concepts}\label{sec:concepts}
We consider $N$ HCB on $d$ lattice sites. The form and dimensionality of the lattice is for the following considerations not relevant.
Let $\mathcal{H}_1^{(d)}$ denote the underlying $d$-dimensional one-particle Hilbert space with an orthonormal basis $\mathcal{B}_1\equiv \{\ket{j}\}_{j=1}^d$ given by the lattice site states $\ket{j}$. Although the `hard-core basis' $\mathcal{B}_1$ might be any basis of one-particle states which, due to some physical constraints, cannot be multiply occupied, we refer in the following to $\ket{j}$ as `sites'.
In case of bosons without hard-core constraint the corresponding $N$-boson Hilbert space $\mathcal{H}_N^{(B)}$ is given by the symmetrized $N$-particle states, namely $\mathcal{H}_N^{(B)}\equiv \mathcal{S}_N \left(\mathcal{H}_1^{(d)}\right)^{\otimes^N}$. Imposing the hard-core constraint then means to restrict this Hilbert space to the subspace $\mathcal{H}_N^{(HCB)}$ of $\mathcal{H}_N^{(B)}$ by excluding configurations with multiply occupied sites. Accordingly, any $N$-HCB state can be expanded as
\begin{equation}\label{Psi}
  \ket{\Psi} = \sum_{\bd{i}}\,A_{\bd{i}}\,\ket{\bd{i}} \,,
\end{equation}
where $\bd{i}\equiv \{i_1,\ldots,i_N\}$, $i_1,\ldots,i_N=1,\ldots,d$, $\ket{\bd{i}}\equiv S_N (\ket{i_1}\otimes \ldots \otimes \ket{i_N})$ is the symmetrization of the $N$-fold tensor product and sums $\sum_{\bd{i}}$ are restricted here and in the following to configurations $\bd{i}$ without multiple occupancies.
It is technically convenient to consider configurations $\bd{i}$ just as \textit{unordered} sets of $N$ (different) indices.
Furthermore, we introduce the corresponding HCB creation, $b_i^\dagger$, and annihilation operators, $b_j$, with respect to the lattice sites. They fulfill mixed commutation relations, i.e.~they commute for different sites and anticommute at the same site \cite{Matsubara1956}.

In contrast to the Hilbert space of $N$ identical fermions or bosons, the $N$-HCB Hilbert space is not invariant under simultaneous one-particle unitary transformations, $ \left(U_1\right)^{\otimes^N}\mathcal{H}_N^{(HCB)} \neq \mathcal{H}_N^{(HCB)}$. The same of course also holds for the algebra of observables: A change of the basis leads to rather odd, namely mixed anticommutation/commutation relations between the new creation, $b_{\alpha}^\dagger$, and annihilation operators $b_\beta$. As a consequence, a possible upper bound on the occupancy $N^{(\varphi)}$ of a  given one-particle state $\ket{\varphi}\in \mathcal{H}_1^{(d)}$ (which can be written as a linear combination of the states $\{\ket{j}\}$) depends highly on $\ket{\varphi}$ itself. Therefore, one-particle states $\ket{\varphi}$ allowing for multiple occupancies may exist, but they need to differ from the lattice site states
$\{\ket{j}\}$.

\section{Maximum occupation number}\label{sec:Nmax}
To determine the optimal universal upper bound on occupation numbers for $N$ HCB on $d$ sites we need to determine
\begin{equation}\label{nmax1}
  N_{max}= \max_{\scriptsize\begin{array}{l}|\varphi\rangle\in \mathcal{H}_1^{(d)} \\ \langle \varphi|\varphi\rangle=1  \end{array}} \max_{\scriptsize\begin{array}{l}|\Psi\rangle\in \mathcal{H}_N^{(HCB)} \\ \langle \Psi|\Psi\rangle=1  \end{array}} \big(N^{(\varphi)}(\ket{\Psi})\big)\,.
  \end{equation}
where  $N^{(\varphi)}(\ket{\Psi})\equiv \langle \Psi| b_\varphi^\dagger b_\varphi|\Psi\rangle$ with $ b_\varphi^\dagger$ and  $b_\varphi$ the creation and annihilation operator of particles in the state $\ket{\varphi}$.
We first present the final result for $N_{max}$ in the form of a theorem and provide its derivation afterwards.
\begin{thm}\label{thm1}
For $N$ hard-core bosons on $d$ lattice sites the maximum possible occupation number is given by
\begin{equation}\label{eq:lambdaMax}
  N_{max}^{(N,d)}\equiv \frac{N}{d}(d-N+1)\,.
  \end{equation}
Only one-particle states $\ket{\varphi_{max}}$ unbiased with respect to the lattice basis $\{\ket{j}\}_{j=1}^{d}$, i.e.~$|\langle j|\varphi_{max}\rangle|^2=\frac{1}{d}$, $\forall j=1,\ldots,d$, allow for such an occupancy, where the corresponding unique and pure maximizer state $\ket{\Psi_{max}}$ follows as
\begin{equation}\label{eq:lambdaMaxState}
  \ket{\Psi_{max}} = \mathcal{N}\,\sum_{\bd{j}} e^{i \phi_{\bd{j}}} \ket{\bd{j}}\,,
\end{equation}
with $\phi_{\bd{j}} = \sum_{m=1}^N \arg(\langle j_m | \varphi_{max} \rangle)$ and $\mathcal{N} =1/\sqrt{\tiny{\binom{d}{N}}}$.
\end{thm}
Theorem \ref{thm1} provides a \textit{universal} upper bound for the Bose-Einstein condensate concentration for HCB on a lattice.
It is worth noting that these results are independent of both, the spatial dimension and the form of the underlying lattice, and of any microscopic details. Whether the ground state of a given hard-core Hamiltonian will exhibit such macroscopic population of a specific state $\ket{\varphi}$ will depend  on its concrete form.
%It is also worth noticing, however, that for every lattice $\mathcal{L}$ there exists a specific one-particle Hamiltonian $\mathcal{H}_1^{(\mathcal{L})}$ such that the ground state of the corresponding system of HCB will attain the bound.
In addition, the possible maximum occupation number $N_{max}$ exhibits a particle-hole symmetry, i.e.~it takes the same value for $N$ and $[d-(N-1)]$ particles, where the latter corresponds to $(N-1)$ holes. In the thermodynamic limit $N,d\rightarrow \infty$ at fixed number density $n\equiv N/d$ the maximal possible degree $n_{max}\equiv N_{max}/N$ of condensation  follows as (this has already been found in \cite{TothBound}, yet by assuming in advance that $\ket{\varphi_{max}}$ is given by the 0-momentum state)
\begin{equation}\label{numax}
  n_{max}(n) = 1-n\,.
\end{equation}

To prove Theorem \ref{thm1} we express $\ket{\varphi}$ with respect to the hard-core basis,
\begin{equation}\label{phi}
  \ket{\varphi}=\sum_{k=1}^d c_k \ket{k}\,,
\end{equation}
where we assume $c_k$ real and non-negative for all $k$ (possible phases of the $c_k$ could be absorbed by the lattice states $\ket{k}$) and we can assume the $N$-HCB state to be pure.
Eq.~(\ref{phi}) together with the expansion (\ref{Psi}) of $\ket{\Psi}$ yields (see Appendix \ref{app:proof1} for technical details)
\begin{eqnarray}\label{nc}
N^{(\varphi)}(\ket{\Psi})
  &=&  \sum_{\bd{i}'}\sum_{k,l=1}^d  A_{\bd{i}'\cup \{k\}}^\ast A_{\bd{i}'\cup \{l\}}  c_k c_l^\ast \nonumber \\
 &=& \sum_{\bd{i}'} \big|\langle\vec{A}^{(\bd{i}')},\vec{c}\,\rangle\big|^2\,.
\end{eqnarray}
Here, the prime should indicate that $\bd{i}'$ is a configuration of $(N-1)$ HCB. The union $\bd{i}'\cup \{k\}$ then means to add a boson in the state $\ket{k}$ to the configuration $\bd{i}'$. For $k \in \bd{i}'$ we have $\bd{i}'\cup \{k\}= \bd{i}'$ (not allowing for double occupancies) and we therefore define $A_{\bd{i}'\cup \{k\}}=0$ whenever $k \in \bd{i}'$. In the last line we introduced the compact notation $\vec{c}\equiv (c_k)_{k=1}^d$, $\vec{A}^{(\bd{i}')} \equiv (A_{\bd{i}'\cup \{k\}})_{k=1}^d$, with $\big(\vec{A}^{(\bd{i}')}\big)_k \equiv 0$ whenever $k \in \bd{i}'$, and $\langle\cdot,\cdot\rangle$ denotes the standard inner product on the $d$-dimensional complex space. Hence, the expression (\ref{nc}) for the one-particle quantity $N^{(\varphi)}(\ket{\Psi})$ is the squared projection of the vector $\vec{c}$ (which characterizes the one-particle state $\ket{\varphi}$) onto the vector $\vec{A}^{(\bd{i}')}$, summed over all configurations $\bd{i}'$ of $(N-1)$ HCB on $d$ sites.

To prove Theorem \ref{thm1} we would need to variationally maximize the occupation number (\ref{nc}) with respect to both, the $N$-HCB state $\ket{\Psi}$ and the one-particle state $\ket{\varphi}$. Since $N^{(\varphi)}(\ket{\Psi})$ is a polynomial of degree four in $\{A_{\bd{i}}\}$, $\{c_k\}$ the corresponding Euler-Lagrange equations are cubic and therefore possibly too difficult to be solved analytically. Even if an analytical solution could be found it would be difficult to verify that the corresponding Hessian is negative definite.
Instead, we choose an elegant approach avoiding any variational equation. This will be achieved by expressing the inner product in the last line of  Eq.~(\ref{nc}) in two different ways
\begin{eqnarray}\label{CS0}
\langle\vec{A}^{(\bd{i}')},\vec{c}\,\rangle
&=& \langle (A_{\bd{i}'\cup \{k\}})_{k=1}^d ,(\chi_{k\not \in \bd{i}'} c_k)_{k=1}^d\,\rangle\nonumber \\
  &=& \langle (A_{\bd{i}'\cup \{k\}} c_k)_{k=1}^d, (\chi_{k\not \in \bd{i}'})_{k=1}^d\,\rangle   \, ,
\end{eqnarray}
where $\chi_{k\not \in \bd{i}'}=1$ if $k\not \in \bd{i}'$ and zero otherwise.  Application of the Cauchy-Schwartz inequality in the spirit of the first and second line of Eq.~(\ref{CS0}) leads to the estimate (see Appendix \ref{app:proof1})
\begin{eqnarray}\label{CS1}
N^{(\varphi)}(\ket{\Psi})
  &\leq & 1+(N-1)\sum_{\bd{i}} \big|A_{\bd{i}}\big|^2 \,\sum_{k \not \in \bd{i}}\big|c_k\,\big|^2
\end{eqnarray}
and
\begin{eqnarray}\label{CS2}
\lefteqn{N^{(\varphi)}(\ket{\Psi})} \\
&\leq& (d-N+1)- (d-N+1)\sum_{\bd{i}} \big|A_{\bd{i}}\big|^2 \sum_{k \not\in \bd{i}}\big|c_k\,\big|^2 \,,\nonumber
\end{eqnarray}
respectively. The pleasant surprise is that the term $\sum_{\bd{i}} \big|A_{\bd{i}}\big|^2 \,\sum_{k \not \in \bd{i}} \big|c_k\,\big|^2$ appears in the final result of estimates (\ref{CS1}), (\ref{CS2}) with different signs. By taking an appropriate linear combination of both estimates it cancels out and one eventually obtains
\begin{equation}
  N^{(\varphi)}(\ket{\Psi})\leq \frac{N}{d}\,(d-N+1)\,.
\end{equation}

This upper bound on $N^{(\varphi)}(\ket{\Psi})$ can be attained only by those $N$-HCB states $\ket{\Psi}$ and one-particle states $\ket{\varphi}$
for which the vectors $\vec{A}^{(\bd{i}')}$, $(\chi_{k\not \in \bd{i}'} c_k)_{k=1}^d$ and $(A_{\bd{i}'\cup\{k\}}c_k)_{k=1}^d$, $(\chi_{k\not \in \bd{i}'})_{k=1}^d$, respectively, are parallel for all $\bd{i}'$. For the case of real and positive $c_k$, this can be achieved only for $c_k\equiv \frac{1}{\sqrt{d}}$ and $A_{\bd{i}}\equiv 1/\sqrt{\binom{d}{N}}$. The case of arbitrary $c_k$-phases, $c_k=e^{i \phi_k} |c_k|$, can be derived from the result of zero-phases by redefining the lattice site states, $\ket{k}\rightarrow e^{i \phi_k}\ket{k}$. This implies $A_{\bd{i}}\rightarrow e^{i \phi_{\bd{i}}}A_{\bd{i}}$ with $\phi_{\bd{i}} \equiv \sum_{m=1}^N \phi_{i_m}$ which completes the proof.

Taking the hard-core condition $(b_j^{\dagger})^2=0$ and the form of $\ket{\varphi_{max}}$ into account it follows $\ket{\Psi_{max}} \propto (b_{\varphi_{max}}^{\dagger})^N \ket{0}$. As a consequence of this product structure, $\ket{\Psi_{max}}$ has zero entanglement. This equivalently means that $\ket{\Psi_{max}}$ contains no additional information beyond that provided by the one-particle reduced density matrix. Indeed, according to Theorem \ref{thm1} $\ket{\Psi_{max}}$ is \emph{uniquely} determined by its one-particle reduced density matrix. A different but even more fascinating connection between maximal condensate concentration and entanglement can be revealed by asking for the maximal possible occupation number $N_{max}^{(\mathcal{L}_A)}$ for a sublattice $\mathcal{L}_A$ of $\mathcal{L}$ with $d_A (<d)$ sites. Generalizing Theorem \ref{thm1} (see Appendix \ref{app:proof2}) we find that  $N_{max}^{(\mathcal{L}_A)}=(d_A+1)^2/4d_A$ and the sublattice $\mathcal{L}_A$ then contains $\overline{N}_A=(d_A+1)/2$ particles.
$\overline{N}_A$ is the number of particles maximizing the expression $N_{max}^{(N_A,d_A)}$ in Theorem \ref{thm1}. The corresponding $N$ HCB quantum state $\ket{\Psi_{max}^{(\mathcal{L}_A)}}$ is given by (the symmetrization of) $\ket{\Psi_{max}}_A\otimes \ket{\overline{N}_B}_B$, where
$\ket{\Psi_{max}}_A$ is the state (\ref{eq:lambdaMaxState}) for $\overline{N}_A$ HCB on $\mathcal{L}_A$ and $\ket{\overline{N}_B}_B$ any state of $\overline{N}_B = N-\overline{N}_A$ HCB on $\mathcal{L}_B$. The structure of the maximizer state $\ket{\Psi_{max}^{(\mathcal{L}_A)}}$ then shows that maximal \emph{local} (i.e.~in $\mathcal{L}_A$) occupation numbers $N_{max}^{(\mathcal{L}_A)}$ can exist \emph{if and only if} the mode entanglement between $\mathcal{L}_A$ and $\mathcal{L}_B$ is minimal (zero). Hence, the entanglement entropy of the mode reduced density operator of $\mathcal{L}\setminus \mathcal{L}_A$ is expected to be reciprocally related to the largest occupation number within $\mathcal{L}_A$.

\section{Physical realization: The `Hubbard star'}\label{sec:Hubbard}
\begin{figure}
   \includegraphics[width=7cm]{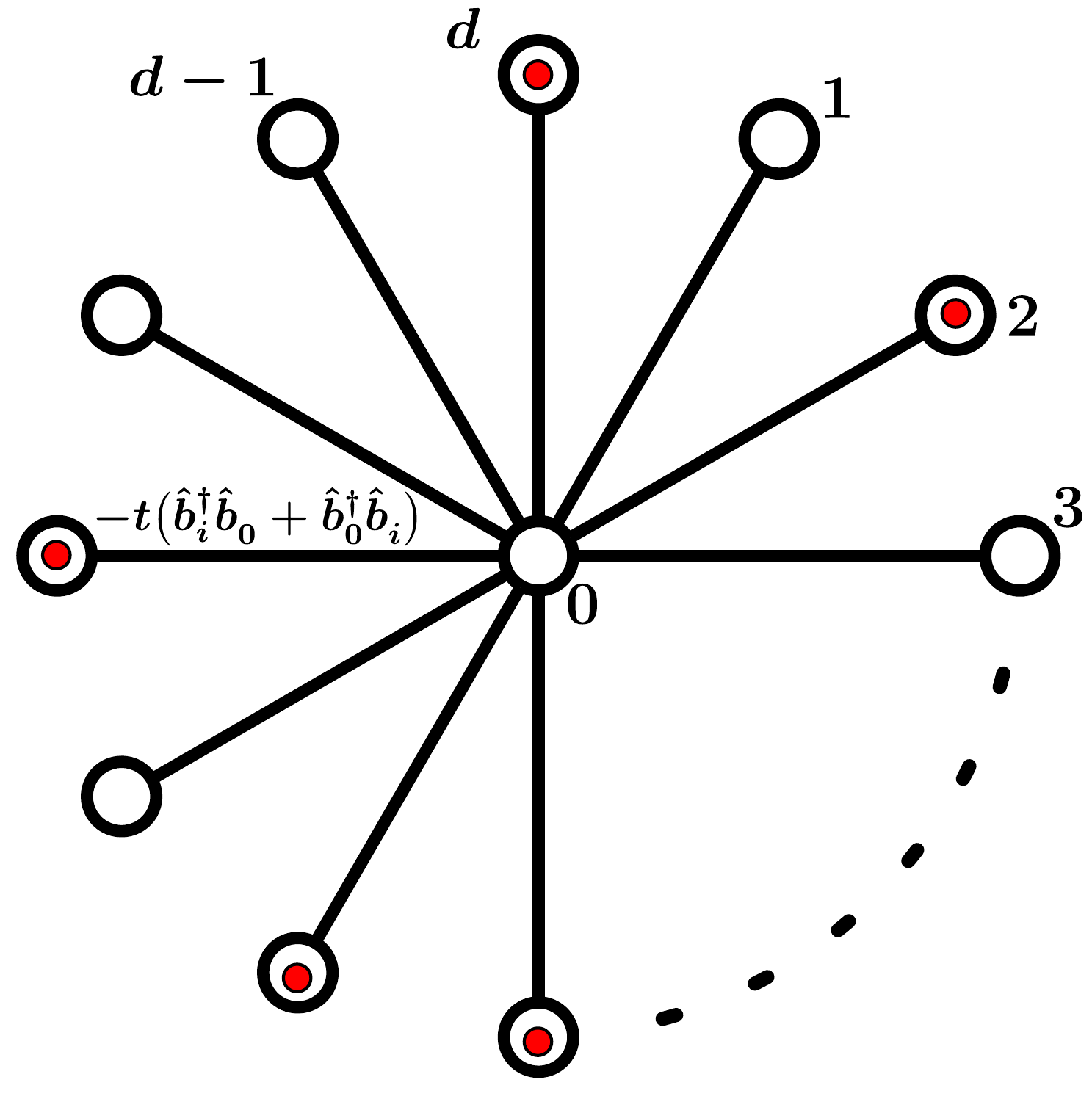}
    \caption[Hubbard Star lattice.]{The (bosonic) Hubbard Star model. Only hopping between the
    outer sites $1$ to $d$ and the central site $0$ is permitted. The open circles represent the
    sites and the full (red) dots the HCB (see text for more details).}
    \label{fig:HubbardStar}
\end{figure}

Concerning the physical relevance of Theorem \ref{thm1} one may wonder whether HCB-Hamiltonians exist having $\ket{\Psi_{max}}$ as ground state. Since all basis states $\ket{j_1,...,j_N}$ contribute equally to $\ket{\Psi_{max}}$, systems with site-independent hopping of the HCB are particulary promising. Indeed, for an infinite-range HCB hopping model without further interactions the ground state is given by $\ket{\Psi_{max}}$ \cite{Toth1990, Penrose1991} (see also Refs.~\cite{Kirson2000, Bru2003, Boland2008}). The experimental realization of such a model, however, seems to be very difficult if not impossible.
We therefore propose here a new model which simulates the infinite-range hopping:
Consider a ring with equally spaced sites $1$ to $d$ and a site $0$ at its center (cf.~Fig.~\ref{fig:HubbardStar}).
We further assume that hopping between different sites on the ring is negligible compared to the hopping between the ring-sites and the central site at a rate of $t >0$. The resulting hard-core Hamiltonian is given by
\begin{equation}\label{eq:HamiltonianHubbardStar}
\hat{H} = -t\sum_{i=1}^{d} b_0^{\dagger} b_i + h.c.~\,.
\end{equation}
Here, $b_j^{\dagger}$ and $b_j$ are the HCB creation and annihilation operators fulfilling the conventional mixed commutation relations \cite{Matsubara1956}.
It is easy to see that $\hat{H}^2$ (describing 2nd order processes) contains hopping terms between \textit{all} ring-sites with identical hopping parameters $t^2$.
$\hat{H}$ conserves the total particle number which allows the restriction to a Hilbert space with fixed particle number $N$. The model shall be called the \textit{(bosonic) Hubbard star} in analogy to the fermionic version studied in Ref.~\cite{vDongen1991}.

The form of Eq.~(\ref{eq:HamiltonianHubbardStar}) makes explicit the connection of HCB to spin systems with spin one half, as already pointed out in Ref. \cite{Matsubara1956}:
According to the Holstein-Primakoff transformation \cite{Holstein1940}, the operators $b_k, b_k^\dagger$ for every site $k$ can be mapped to spin operators for a spin $1/2$ (with $\hbar\equiv 1$)
\begin{eqnarray}\label{spins}
S_k^+ &\equiv &\sqrt{1-b_k^\dagger b_k}\,\, b_k\,,\,\,\,S_k^-\equiv \left(S_k^+\right)^\dagger=b_k^\dagger \sqrt{1-b_k^\dagger b_k} \nonumber \\
S_k^z&\equiv& \frac{1}{2}-b_k^\dagger b_k\,.
\end{eqnarray}
Here, $S_k^\pm$ are the corresponding spin ladder operators and the original bosonic vacuum state $\ket{0}$ is mapped to the completely polarized spin state $\ket{\!\uparrow}_0 \otimes \ket{\!\uparrow, \ldots,\uparrow}_R$. It is straightforward to verify that the operators in (\ref{spins}) fulfill the commutation relations for spin $1/2$.
The Holstein-Primakoff transformation then maps the Hamiltonian (\ref{eq:HamiltonianHubbardStar}) to the corresponding spin model
\begin{equation}\label{eq:HamiltonianHubbardStarRingSpin}
\hat{H}' = -t\big(\hat{S}^{+}_{0} \hat{S}^{-}_R + \hat{S}^{-}_0 \hat{S}^{+}_R\big) \,,
\end{equation}
where $\hat{\boldsymbol{S}}_R = \sum_{i=1}^{d} \hat{\boldsymbol{S}}_i$ denotes  the total spin operator on the ring.
Since the creation of a HCB corresponds to a spin flip, $N$-particle states are mapped to states with total magnetic quantum number $M=(d+1)/2-N$.

The eigenstates of Hamiltonian (\ref{eq:HamiltonianHubbardStarRingSpin}) can be expanded as
\begin{equation}\label{eq:AnsatzHubbardStar}
\ket{\psi} = \alpha_1 \ket{\!\uparrow}_0\otimes \ket{\psi_1}_R + \alpha_2 \ket{\!\downarrow}_0\otimes \ket{\psi_2}_R  \ .
\end{equation}
The ring states $\ket{\psi_1}_R$ and $\ket{\psi_2}_R$ are normalized and orthogonal with magnetization $M_1=M-\frac{1}{2}$ and $M_2=M+\frac{1}{2}$, respectively.
The eigenvalue equation $\hat{H}' \ket{\psi}=E \ket{\psi}$ reduces to
\begin{equation}\label{eq:SchroedingerEquationHubbardStar1}
\begin{aligned}
E \alpha_1 \ket{\psi_1}_R &= -t \alpha_2 \hat{S}^{-}_R \ket{\psi_2}_R \\
E \alpha_2 \ket{\psi_2}_R &= -t \alpha_1 \hat{S}^{+}_R \ket{\psi_1}_R \,.
\end{aligned}
\end{equation}
Let $\ket{S_R,M_R}$ be an eigenstate of  $\hat{\boldsymbol{S}}^2_R$ and $\hat{S}^{z}_R$ with eigenvalue $S_R(S_R+1)$ and $M_R$, respectively. By making use of
\begin{eqnarray}\label{eq:EigenEquationSpinRoperator}
\lefteqn{\hat{S}^{+}_{R} \hat{S}^{-}_{R} \ket{S_R,M_R}} \quad \quad&& \nonumber\\
&=& \big[(S_R(S_R+1) - M_R(M_R-1)\big] \ket{S_R,M_R}\,,
\end{eqnarray}
Eq.~(\ref{eq:SchroedingerEquationHubbardStar1}) can easily be solved. With $M_R = M_2$ the ground state eigenvalue follows for maximal $S_R$, $S_R = d/2$,
\begin{equation}
E = - t \sqrt{N(d-N+1)}
\end{equation}
and up to a normalization factor we find
\begin{equation}\label{eq:EigenstateSrMr}
\begin{aligned}
\ket{\psi_1}_R &\propto  \big(\hat{S}^{-}_R\big)^{N} \ket{\!\uparrow,\ldots,\uparrow}_R\\
\ket{\psi_2}_R &\propto \big(\hat{S}^{-}_R\big)^{(N-1)} \ket{\!\uparrow,\ldots,\uparrow}_R  \,.
\end{aligned}
\end{equation}
Substitution into Eq.~\eqref{eq:SchroedingerEquationHubbardStar1} allows one to determine the coefficients $\alpha_i$. Use of the inverse Holstein-Primakoff transformation finally yields the ground state of the $N$ HCB,
\begin{equation}\label{eq:gs}
\begin{aligned}
\ket{\psi} =& \frac{1}{\sqrt{2}} \ket{\Psi_{max}^{(N)}}_R + \frac{1}{\sqrt{2}} \hat{b}_0^{\dagger}\ket{\Psi_{max}^{(N-1)}}_R \,,
\end{aligned}
\end{equation}
Here, $\ket{\Psi_{max}^{(N)}}_R$ denotes the state of maximal occupation number \eqref{eq:lambdaMaxState} of $N$ HCB on $d$ sites of the ring, where the corresponding $\ket{\varphi_{max}}$ is given by the 0-momentum state on the ring (i.e.~$\phi_{\bd{j}}\equiv0$).

Since $\ket{\psi}$ involves the maximizing state $\ket{\Psi_{max}}_R$ of Theorem \ref{thm1} for $N$ and $N-1$ particles on the ring, the ground state $\ket{\psi}$ obviously exhibits fractional BEC. To confirm this also by quantitative means we follow Ref.~\cite{Penrose1956} and calculate the largest eigenvalue of the corresponding one-particle reduced density operator
\begin{equation}\label{eq:1RDO}
\rho_1\equiv N \mbox{Tr}_{N-1}[\ket{\psi}\!\bra{\psi}]\equiv \sum_{j=1}^{d+1}\lambda_j \ket{\chi_j}\!\bra{\chi_j}\,,
\end{equation}
obtained by tracing out $N-1$ HCB.
In particular, we determine its eigenstates (natural orbitals $\ket{\chi_j}$) and eigenvalues (natural occupation numbers $\lambda_j$).
Since $\ket{\psi}$ is invariant under arbitrary permutations of the ring sites this is straightforward: Let $U(\pi)$ be an arbitrary permutation of the ring site states, $U(\pi)\ket{j}=\ket{\pi(j)}, j=1,2,\ldots,d$, where the central site state is not affected, $U(\pi)\ket{0}=\ket{0}$. Then, the structure of the ground state (\ref{eq:gs}) (recall also Theorem \ref{thm1}) implies for all $\pi$  $U(\pi)^{\otimes^N}\ket{\psi}=\ket{\psi}$. Since $U(\pi)$ is a unitary operator, the one-particle reduced density operator (\ref{eq:1RDO}) inherits that symmetry, i.e.~one has
\begin{equation}\label{eq:1RDOsym}
[\rho_1,U(\pi)]=0\,,\quad \forall \pi\,.
\end{equation}
As a consequence, $\rho_1$ is block-diagonal with respect to the eigenspaces of \emph{all} $U(\pi)$. Moreover, we observe that only the two states $\ket{0}$ and $1/\sqrt{d}\sum_{j=1}^d \ket{j}$ (and their linear combinations) are eigenstates of all $U(\pi)$ (always with eigenvalue $1$). The $(d-1)$-dimensional subspace $\mathcal{H}_2^\perp$ orthogonal to the space $\mathcal{H}_2$ spanned by those two states is therefore an irreducible representation of the group of ring site permutations. Thus, $d-1$ natural occupation numbers $\lambda_j$ are degenerate and their respective natural orbitals $\ket{\chi_j}$ span the space $\mathcal{H}_2^\perp$. To determine the remaining two natural orbitals and natural occupation numbers we express $\rho_1$, restricted to $\mathcal{H}_2$, with respect to the states $\ket{0}$, $1/\sqrt{d}\sum_{j=1}^d \ket{j}$, leading to
\begin{equation}\label{eq:NON2dim}
\rho_1|_{\mathcal{H}_2} = \frac{1}{2} \left(\begin{array}{cc} 1& \sqrt{N_{max}^{(N,d)}}\\ \sqrt{N_{max}^{(N,d)}} & N_{max}^{(N,d)}+ N_{max}^{(N-1,d)}\end{array}\right)\,.
%&=& \frac{1}{2} \left(\begin{array}{cc} 1& \sqrt{N_{max}^{(N,d)}}\\ \sqrt{N_{max}^{(N,d)}} & 2 N_{max}^{(N,d)}-\left(1-\frac{2(N-1)}{d}\right)\end{array}\right)
\end{equation}
The matrix (\ref{eq:NON2dim}) can easily be diagonalized, leading to the remaining two natural orbitals and natural occupation numbers (the concrete value of the other $d-1$ (degenerate) NON can then be determined via the normalization of $\rho_1$). We state the concrete results for the thermodynamic limit, i.e.~$N,d\rightarrow \infty$ at fixed filling factor $n\equiv N/(d+1)$. The two eigenvalues of (\ref{eq:NON2dim}) in leading order follow as $N (1-n)$ and $1/4$. The normalization of $\rho_1$ then implies that the other eigenvalues in leading order are given by $n^2$, i.e.~they are not macroscopic in $N$. This result shows that BEC is present with the maximal possible degree $n_{max}\equiv 1-n$ of condensation (recall Eq.~(\ref{numax})). The respective  maximally occupied one-particle state is given in (leading order) by the 0-momentum state on the ring, i.e.~$\ket{\varphi_{max}}=1/\sqrt{d}\sum_{j=1}^d \ket{j}$

For the sake of completeness, we mention another model which has $\ket{\Psi_{max}}$ as its ground state. It is a one-dimensional lattice gas model with nearest neighbour hopping and nearest neighbour interactions, provided the ratio of the hopping parameter and the coupling constant takes a very \textit{specific} value \cite{Yang1966}. The precise tuning of the coupling constant may be again difficult in practice.

\section{Summary and conclusions}\label{sec:concl}
For $N$ hard-core bosons on a lattice of $d$ sites we have proven that the hard-core constraint itself enforces a non-trivial universal upper bound on arbitrary occupation numbers. The maximal possible occupation number $N_{max}=(N/d)(d-N+1)$ is proportional to the relative `free volume' $(d-N+1)/d$, i.e., to the density of available sites. This upper bound $N_{max}$ can be attained only for one-particle states $\ket{\varphi_{max}}$ which are maximally unbiased with respect to the hard-core basis (sites). The corresponding unique and pure $N$-HCB maximizer state $\ket{\Psi_{max}}$ is maximally delocalized (cf.~Theorem \ref{thm1}).
Accordingly, $\ket{\varphi_{max}}$ corresponds to a one-particle state with zero `momentum', which has a macroscopic occupancy in the state  $\ket{\Psi_{max}}$.
Since all these results are independent of the spatial dimension and form of the underlying lattice and of the microscopic details and temperature of the system, our work establishes a new, much broader perspective on BEC: It is based on the structure of the $N$-HCB state space only and does not refer to properties of some specific Hamiltonians. In addition, from a general viewpoint, we have also shown that (incomplete) BEC is possible for every lattice despite the hard-core repulsion.

The significance of our universal result \ref{thm1} has been confirmed through the existence of two well-known models whose ground states exhibit the maximal possible degree of condensation. One of them, the infinite-range hopping model for ``free'' HCB also shows that the largest occupation number is strongly related to the mobility of the HCB.
The fact that the infinite-range hopping model attains the upper bound $N_{max}$ is not surprising due to the mean-field character of that model. Indeed, it is known that the order parameter given by the `degree of condensation' becomes maximal in mean-field approximations.
Since its experimental realization, however, is very difficult if not impossible we have proposed in the form the Hubbard star a new HCB model which simulates the infinite-range hopping.
The experimental realization of the Hubbard star exhibiting BEC of maximal degree seems to be feasible. Indeed, the experimentalists in the field of ultra-cold gases have demonstrated high skills by realizing various models (see, e.g., Refs.~\cite{Bloch2002,Paredes2004,Bloch2008,JochimHubb}).
By generating a ring-like optical lattice including a central potential well and by tuning the barrier heights in order to make the hopping between the central and the ring-sites dominant our predictions can be tested. It is also worth noting that it is the \textit{single} site at the ring's center which makes BEC possible by drastically increasing the mobility of the HCB on the ring.
In case that the ring hopping parameter vanishes, $t_{R}=0$, it is the central site only which generates an effective mobility (via 2nd order processes) on the ring.  In an experiment, it would be therefore interesting to increase the ratio $t_{R}/t$ more and more.
For values much smaller than unity nothing will change qualitatively due to the gap in the spectrum of the Hubbard star Hamiltonian (\ref{eq:HamiltonianHubbardStar}). However, at $t_{R}/t = \mathcal{O}(1)$ there will be a crossover from a condensate with $N_{max}(N) \propto N$ to  $N_{max}(N) \propto \sqrt{N}$ (c.f \cite{Schultz1963, Lenard1964}).

Our results also reveal an interesting link between BEC and entanglement: The maximum possible condensate concentration for HCB on a lattice $\mathcal{L}$, or on a sublattice $\mathcal{L}_A$, occurs for states with zero entanglement. This observation adds a new facet to BEC by building a bridge to quantum information theory. Moreover, in the same context, our work could be understood as a first step towards addressing the famous and fundamentally important one-body $N$-representability problem \cite{Coulson} for HCB: Calculating \emph{all} constraints on the one-particle picture emerging from the mixed HCB commutation relations could lead to new insights into, e.g., quantum pase transitions in systems of HCB.

\begin{acknowledgements}
We thank P.G.J.\hspace{0.5mm}van Dongen, M.\hspace{0.5mm}Rizzi and M.\hspace{0.5mm}Streif for helpful discussions.
We gratefully acknowledge financial support from
the Friedrich-Naumann-Stiftung and Christ Church Oxford (FT), the Oxford Martin School, the NRF (Singapore), the MoE (Singapore) and the EU Collaborative Project TherMiQ (Grant Agreement 618074)
(VV), the Oxford Martin Programme on Bio-Inspired Quantum Technologies and the UK Engineering  and Physical Sciences Research Council (Grant EP/P007155/1) (CS).
\end{acknowledgements}

\appendix

\section{Proof of Theorem 1}\label{app:proof1}
We consider the expectation value of the occupancy, $N^{(\varphi)}(\ket{\Psi})$, of $\ket{\varphi}$ (\ref{phi}) given that the system of $N$ HCB is in the state $\ket{\Psi}$ (\ref{Psi}). We derive a compact expression for this quantity:
\begin{eqnarray}\label{app:nc}
N^{(\varphi)}(\ket{\Psi})
  &\equiv & \bra{\Psi} b_{\varphi}^\dagger b_{\varphi} \ket{\Psi}\nonumber \\
  &=& \sum_{\bd{i},\bd{j}}  A_{\bd{i}}^\ast A_{\bd{j}} \sum_{k,l=1}^d c_k c_l^\ast\bra{\bd{i}}b_k^\dagger b_l \ket{\bd{j}}\nonumber \\
  &=& \sum_{\bd{i},\bd{j}}  A_{\bd{i}}^\ast A_{\bd{j}} \sum_{k \in \bd{i},l \in \bd{j}} c_k c_l^\ast \bra{\bd{i}}b_k^\dagger b_l \ket{\bd{j}}\nonumber \\
  &=& \sum_{\bd{i},\bd{j}}  A_{\bd{i}}^\ast A_{\bd{j}} \sum_{k \in \bd{i},l \in \bd{j}} c_k c_l^\ast ´\delta_{\bd{i}\setminus\{k\},\bd{j}\setminus\{l\}}\nonumber \\
  &=&  \sum_{\bd{i}'}\sum_{k,l=1}^d  A_{\bd{i}'\cup \{k\}}^\ast A_{\bd{i}'\cup \{l\}}  c_k c_l^\ast \nonumber \\
  &=& \sum_{\bd{i}'} \left(\sum_{k=1}^d  A_{\bd{i}'\cup \{k\}}^\ast c_k\right)\,\left(\sum_{l=1}^d  A_{\bd{i}'\cup \{l\}}  c_l^\ast\right) \nonumber \\
 &=& \sum_{\bd{i}'} \big|\langle\vec{A}^{(\bd{i}')},\vec{c}\,\rangle\big|^2\,.
\end{eqnarray}
In line four, $\delta$ denotes the Kronecker delta. The prime should indicate that $\bd{i}'$ is a configuration of $(N-1)$ HCB (in contrast to $\bd{i}$ being a configuration of $N$ HCB). The union $\bd{i}'\cup \{k\}$ then means to add a boson in the state $\ket{k}$ to the configuration $\bd{i}'$. For $k \in \bd{i}'$ we have $\bd{i}'\cup \{k\}= \bd{i}'$ (not allowing for double occupancies) and we therefore define $A_{\bd{i}'\cup \{k\}}=0$ whenever $k \in \bd{i}'$. In the last line we introduced the compact notation $\vec{c}\equiv (c_k)_{k=1}^d$, $\vec{A}^{(\bd{i}')} \equiv (A_{\bd{i}'\cup \{k\}})_{k=1}^d$, with the $k$-th component, $\big(\vec{A}^{(\bd{i}')}\big)_k \equiv 0$ whenever $k \in \bd{i}'$, and $\langle\cdot,\cdot\rangle$ denotes the standard inner product on $d$-dimensional complex space. Hence, the expression (\ref{app:nc}) for the one-particle quantity $N^{(\varphi)}(\ket{\Psi})$ is the squared projection of the vector $\vec{c}$ (which characterizes the one-particle state $\ket{\varphi}$) onto the vector $\vec{A}^{(\bd{i}')}$, summed over all configurations
$\bd{i}'$ of $(N-1)$ HCB on $d$ sites.

Since $N^{(\varphi)}(\ket{\Psi})$ is a polynomial of degree four in $\{A_{\bd{i}}\}$, $\{c_k\}$ the corresponding Euler-Lagrange equations are cubic and therefore possibly too difficult to solve analytically. Instead, we choose an elegant approach avoiding  any variational equation. This will be achieved by expressing the inner product in the last line of  Eq.~(\ref{app:nc}) in two different ways
\begin{eqnarray}\label{app:CS0}
\langle\vec{A}^{(\bd{i}')},\vec{c}\,\rangle
&=& \langle (A_{\bd{i}'\cup \{k\}})_{k=1}^d ,(\chi_{k\not \in \bd{i}'} c_k)_{k=1}^d\,\rangle\nonumber \\
  &=& \langle (A_{\bd{i}'\cup \{k\}} c_k)_{k=1}^d, (\chi_{k\not \in \bd{i}'})_{k=1}^d\,\rangle   \, ,
\end{eqnarray}
where $\chi_{k\not \in \bd{i}'}=1$ if $k\not \in \bd{i}'$ and zero otherwise. Application of the Cauchy-Schwartz inequality in the spirit of the first line of Eq.~(\ref{app:CS0}) yields for (\ref{app:nc})
\begin{eqnarray}\label{app:CS1}
N^{(\varphi)}(\ket{\Psi})
 &=& \sum_{\bd{i}'} \big|\langle \vec{A}^{(\bd{i}')} ,(\chi_{k\not \in \bd{i}'} c_k)_{k=1}^d\,\rangle\big|^2\nonumber \\
  &\leq & \sum_{\bd{i}'} \big|\vec{A}^{(\bd{i}')}\big|^2 \,\, \big|(\chi_{k\not \in \bd{i}'} c_k)_{k=1}^d\big|^2\nonumber \\
  &= & \sum_{\bd{i}'} \Big(\sum_{l=1}^d\big|A_{\bd{i}'\cup\{l\}}\big|^2\Big) \, \sum_{k \not \in \bd{i}'}\big|c_k\,\big|^2\nonumber \\
  &= & \sum_{l=1}^d\sum_{\bd{i} \ni l} \big|A_{\bd{i}}\big|^2 \, \sum_{k\not \in (\bd{i}\setminus \{l\})}\big|c_k\,\big|^2\nonumber \\
  &=& \sum_{\bd{i}} \big|A_{\bd{i}}\big|^2 \,\sum_{l \in \bd{i}}\,\sum_{k\not \in (\bd{i}\setminus \{l\})}\big|c_k\,\big|^2\nonumber \\
  &=& \sum_{\bd{i}} \big|A_{\bd{i}}\big|^2 \,\sum_{l \in \bd{i}}\,\Big(\sum_{k\not \in \bd{i}}\big|c_k\,\big|^2+\big|c_l\,\big|^2\Big)\nonumber \\
  &=& \sum_{\bd{i}} \big|A_{\bd{i}}\big|^2 \,\big[N \sum_{k \not \in \bd{i}}\big|c_k\,\big|^2+\sum_{l \in \bd{i}}\big|c_l\,\big|^2\big]\nonumber \\
  &=& 1+(N-1)\sum_{\bd{i}} \big|A_{\bd{i}}\big|^2 \,\sum_{k \not \in \bd{i}}\big|c_k\,\big|^2 \,.
\end{eqnarray}
In the forth line, $\sum_{\bd{i} \ni l}$ denotes the sum over those configurations $\bd{i}$ of $N$ HCB which contain the site index $l$ and $\bd{i}'$ can be written as $\bd{i}\setminus \{l\}$. In the last line we have first used the normalization of $\ket{\varphi}$ and then of $\ket{\Psi}$. Application of the Cauchy-Schwartz inequality in the spirit of the second line of Eq.~(\ref{CS0}) yields for (\ref{nc})
\begin{eqnarray}\label{app:CS2}
N^{(\varphi)}(\ket{\Psi})
&=& \sum_{\bd{i}'} \big|\langle (A_{\bd{i}'\cup\{k\}}c_k)_{k=1}^d ,(\chi_{k\not \in \bd{i}'})_{k=1}^d\,\rangle\big|^2 \nonumber \\
  &\leq & \sum_{\bd{i}'} \big|(A_{\bd{i}'\cup\{k\}}c_k)_{k=1}^d\big|^2 \,\big|(\chi_{k\not \in \bd{i}'})_{k=1}^d\big|^2 \nonumber \\
  &=& (d-N+1)\sum_{\bd{i}} \big|A_{\bd{i}}\big|^2 \sum_{k \in \bd{i}}\big|c_k\,\big|^2  \\
  &=& (d-N+1)- (d-N+1)\sum_{\bd{i}} \big|A_{\bd{i}}\big|^2 \sum_{k \not\in \bd{i}}\big|c_k\,\big|^2 \,.\nonumber
\end{eqnarray}
In the third line, we have used $\big|(\chi_{k\not \in \bd{i}'})_{k=1}^d\big|^2=d-N+1$ for all $\bd{i}'$ and for $k \not \in \bd{i}'$ we introduced $\bd{i}=\bd{i}'\cup\{k\}$. In the forth line, we have first used the normalization of $\ket{\varphi}$ and then of $\ket{\Psi}$.

The pleasant surprise is that the term $\sum_{\bd{i}} \big|A_{\bd{i}}\big|^2 \,\sum_{k \not \in \bd{i}} \big|c_k\,\big|^2$ appears in the final result of estimates (\ref{app:CS1}), (\ref{app:CS2}) with different signs. By considering the specific linear combination
$(d-N+1)(\mbox{A3}) +(N-1)(\mbox{A4})$ of estimate (\ref{app:CS1}) and (\ref{app:CS2}) it cancels out:
\begin{eqnarray}
\lefteqn{(d-N+1)\,N^{(\varphi)}(\ket{\Psi}) +(N-1)\,N^{(\varphi)}(\ket{\Psi})}\qquad \nonumber \\
&\leq& (d-N+1)+(d-N+1)(N-1)\nonumber \\
&=&N (d-N+1)\,.
\end{eqnarray}
Eventually, this leads to
\begin{equation}
  N^{(\varphi)}(\ket{\Psi})\leq \frac{N}{d}\,(d-N+1)\,.
\end{equation}

This upper bound on $N^{(\varphi)}(\ket{\Psi})$ can be attained only by those $N$-HCB states $\ket{\Psi}$ and one-particle states $\ket{\varphi}$
for which the vectors $\vec{A}^{(\bd{i}')}$, $(\chi_{k\not \in \bd{i}'} c_k)_{k=1}^d$ and $(A_{\bd{i}'\cup\{k\}}c_k)_{k=1}^d$, $(\chi_{k\not \in \bd{i}'})_{k=1}^d$, respectively, are parallel for all $\bd{i}'$. For the case of $c_k \in \R_0^+, \forall k$, this can be achieved only for $c_k\equiv \frac{1}{\sqrt{d}}$ and $A_{\bd{i}}\equiv 1/\sqrt{\binom{d}{N}}$. The case of arbitrary $c_k$-phases, $c_k=e^{i \phi_k} |c_k|$, can be derived from the result of zero-phases by redefining the lattice site states, $\ket{k}\rightarrow e^{i \phi_k}\ket{k}$. This implies $A_{\bd{i}}\rightarrow e^{i \phi_{\bd{i}}}A_{\bd{i}}$ with $\phi_{\bd{i}} \equiv \sum_{m=1}^N \phi_{i_m}$ which completes the proof.

\section{A generalized theorem and its proof}\label{app:proof2}
From a practical viewpoint, particularly for macroscopically large lattice systems $\mathcal{L}$ the concept of a subsystem plays an important role and a natural question arises: What is the maximal possible occupation number that one can find \emph{within} a subsystem $\mathcal{L}_A$ of $d_A<d$ sites? The answer to this important question is given by the following theorem:

\begin{thm}\label{thm2}
For $N$ hard-core bosons on a lattice $\mathcal{L}$ of $d$ sites the maximum possible occupation number that can be found within a sublattice $\mathcal{L}_A$ of $d_A$ sites is given by
\begin{equation}\label{eq:lambdaMax2}
  N_{max}^{(\mathcal{L}_A)}\equiv \max_{N_A^{-}\leq N_A \leq N_A^{+}}\big[N_{max}^{(N_A,d_A)}\big]\,.
  \end{equation}
where $N_A^{-}=\max{\big(0,N-(d-d_A)\big)}$, $N_A^{+}=\min{(N,d_A)}$ and $N_{max}$ is given by (\ref{eq:lambdaMax}).
Only one-particle states $\ket{\varphi_{max}^{(\mathcal{L}_A)}}$ unbiased with respect to the lattice states $\{\ket{j}\}_{j\in \mathcal{L}_A}$ of the sublattice $\mathcal{L}_A$ allow for such an occupancy. Any  maximizer state $\ket{\Psi_{max}^{(\mathcal{L}_A)}}$ has the form
\begin{equation}\label{PsiABmax}
\ket{\Psi_{max}^{(\mathcal{L}_A)}} = \mathcal{S}_N\left[\ket{\Psi_{max}}_A\otimes \ket{N-\overline{N}_A}_B\right]\,,
\end{equation}
where $\overline{N}_A$ is the particle number maximizing (\ref{eq:lambdaMax2}), $\ket{\Psi_{max}}_A$ the maximizer state for $\overline{N}_A$ HCB on the sublattice $\mathcal{L}_A$ of $d_A$ sites according to Theorem 1, $\ket{N-\overline{N}_A}_B$ an arbitrary state of $N-\overline{N}_A$ HCB on the complementary lattice $\mathcal{L}\setminus \mathcal{L}_A$ and $\mathcal{S}_N$ denotes the symmetrizing operator for $N$ particles.
\end{thm}

Let us label the $d$ lattice sites of the total lattice $\mathcal{L}$ such that the sites $1,2,\ldots,d_A$ belong to the sublattice $\mathcal{L}_A$. The sites of its complementary lattice $\mathcal{L}_B \equiv \mathcal{L}\setminus\mathcal{L}_A$ are then labeled by $d_A+1,\ldots,d$. The one-particle Hilbert space $\mathcal{H}_1(\mathcal{L})$ for the total lattice splits according to
\begin{equation}
\mathcal{H}_1(\mathcal{L})= \mathcal{H}_1(\mathcal{L}_A)\oplus \mathcal{H}_1(\mathcal{L}_B)
\end{equation}
since any one-particle quantum state $\ket{\varphi} \in \mathcal{H}_1(\mathcal{L})$ is expressed in a unique way as $\ket{\varphi}=\sum_{k=1}^{d_A} c_k \ket{k} + \sum_{k=d_A+1}^{d} c_k \ket{k}$. This structure of the one-particle Hilbert space implies that the corresponding HCB Fock space $\mathcal{F}^{(HCB)}$ over $\mathcal{H}_1(\mathcal{L})$ has the following structure
\begin{equation}\label{Fisom}
\mathcal{F}^{(HCB)} \cong \mathcal{F}^{(HCB)}_A \otimes \mathcal{F}^{(HCB)}_B\,,
\end{equation}
where $\mathcal{F}^{(HCB)}_{A/B}$ denote the respective HCB Fock spaces over $\mathcal{H}_1(\mathcal{L}_{A/B})$.
The isomorphism (\ref{Fisom}) is rather elementary. It is given by
\begin{equation}
b_{j_1}^\dagger\cdot\ldots \cdot b_{j_N}^\dagger\ket{0} \leftrightarrow \Big(\prod_{j_i\leq d_A} b_{j_i}^\dagger \ket{0}_A\Big) \otimes \Big(\prod_{j_i>  d_A} b_{j_i}^\dagger \ket{0}_B\Big)\,,
\end{equation}
for all $N=0,1,\ldots,d$, and for all sets of different $j_1,\ldots,j_N \in \{1,2,\ldots,d\}$, where we used again second quantization and introduced the vacuum states for $\mathcal{F}^{(HCB)}$ ($\ket{0}$), $\mathcal{F}^{(HCB)}_A$ ($\ket{0}_A$) and $\mathcal{F}^{(HCB)}_B$ ($\ket{0}_B$).

We use in the following the expansion
%now assume that the maximal possible occupation number within $\mathcal{L}_A$ is attained for the one particle quantum state
\begin{equation}\label{phiA}
\ket{\varphi} = \sum_{k=1}^{d_A} c_k \ket{k}
\end{equation}
and
%with corresponding $N$-HCB maximizer state $\ket{\Psi} \in \mathcal{F}^{(HCB)}$. We can express $\ket{\Psi}$ in a compact way as
\begin{equation}\label{PsiAB}
\ket{\Psi} = \sum_{\bd{i}} A_{\bd{i}} \ket{\bd{i}} = \sum_{\bd{i}_A, \bd{i}_B} A_{\bd{i}_A\cup \bd{i}_B} \ket{\bd{i}_A\cup \bd{i}_B}\,.
\end{equation}
Here, the sum $\sum_{\bd{i}}$ contains all configuration of $N$ HCB on $d$ sites. The sum $\sum_{\bd{i}_A}$ and $\sum_{\bd{i}_B}$ denote sums over configurations within the lattice $\mathcal{L}_A$ and $\mathcal{L}_B$, respectively. Since the latter two sums are not restricted to a fixed particle number we need to define $A_{\bd{i}_A\cup \bd{i}_B}\equiv0$ whenever $\bd{i}_A\cup \bd{i}_B$ is not a configuration of $N$ HCB. We can now begin to calculate the corresponding particle number expectation value.

\begin{eqnarray}\label{NA1}
N^{(\varphi)}(\ket{\Psi}) &\equiv& \bra{\Psi}b_{\varphi}^\dagger b_{\varphi}\ket{\Psi}  \nonumber \\
&=& \mbox{Tr}_{\mathcal{F}^{(HCB)}}\big[b_{\varphi}^\dagger b_{\varphi} \ket{\Psi}\!\bra{\Psi}\big] \nonumber \\
&=& \mbox{Tr}_{\mathcal{F}^{(HCB)}_A}\big[b_{\varphi}^\dagger b_{\varphi} \rho_A\big]   \,,
\end{eqnarray}
where we introduced the mode-reduced density operator, $\rho_A \equiv \mbox{Tr}_{\mathcal{F}_B^{(HCB)}}[\ket{\Psi}\!\bra{\Psi}]$, for subsystem $\mathcal{L}_A$ and made use of the fact that $\ket{\varphi}$ contains only lattice sites of system $\mathcal{L}_A$. Since the state $\ket{\Psi}$ for the total system has fixed particle number, the reduced state $\rho_A$ is block-diagonal with respect to the different particle number sectors. By introducing the operator $\hat{P}_A^{(N_A)}$ projecting $\mathcal{F}^{(HCB)}_A$ onto the subspace of fixed particle number $N_A$ we have $\rho_A= \sum_{N_A=0}^N\,\hat{P}_A^{(N_A)} \rho_A \hat{P}_A^{(N_A)}$. Depending on the concrete values of $N,d$ and $d_A$ it is possible to further restrict
this sum since not all particle numbers $N_A$ between $0$ and $N$ are possible on $\mathcal{L}_A$. For instance, for the case $N=d-1$ and $d_A=d-1$ only particle numbers $N_A=N-1,N$ are possible. In general, the sum can be restricted to the minimal ($N_A^{-}$) and maximal possible particle number ($N_A^{+}$) following as
\begin{equation}\label{NAbounds}
N_A^{-}=\max{\big(0,N-(d-d_A)\big)}\,,\,\,\,N_A^{+}=\min{(N,d_A)}\,.
\end{equation}
Consequently, we can express $\rho_A$ as
\begin{equation}\label{rhoANA}
\rho_A = \sum_{N_A=N_A^{-}}^{N_A^{+}} q^{(N_A)} \rho_A^{(N_A)}\,,
\end{equation}
where the state $\rho_A^{(N_A)}$ has particle number $N_A$ and is trace-normalized to one. Hence, we have
\begin{eqnarray}\label{qNA}
q^{(N_A)}&\equiv&\mbox{Tr}_{\mathcal{F}^{(HCB)}_A}[\hat{P}_A^{(N_A)} \rho_A \hat{P}_A^{(N_A)}] \nonumber \\
 &=& \sum_{|\bd{i}_A|=N_A}\,\, \sum_{|\bd{i}_B|=N-N_A} \big|A_{\bd{i}_A\cup \bd{i}_B}\big|^2\,,
\end{eqnarray}
where $\sum_{|\bd{i}_A|=N_A}$ denotes the sum over all configurations $\bd{i}_A$ on $\mathcal{L}_A$ with particle number $|\bd{i}_A|=N_A$ (and analogously $\sum_{|\bd{i}_B|=N_B}$). In principle, one could also restrict the trace over $_{\mathcal{F}^{(HCB)}_A}$ in Eq.~(\ref{qNA}) to the particle number sector with $N_A$ particles. Plugin in the expression (\ref{rhoANA}) in Eq.~(\ref{NA1}) yields
\begin{eqnarray}\label{NA2}
N^{(\varphi)}(\ket{\Psi}) = \sum_{N_A=N_A^{-}}^{N_A^{+}} q^{(N_A)} \mbox{Tr}_{\mathcal{F}^{(HCB)}_A}\big[b_{\varphi}^\dagger b_{\varphi} \rho_A^{(N_A)}\big]  \,.
\end{eqnarray}
The crucial point is now that $N^{(\varphi)}(\ket{\Psi})$ is a convex combination (indeed we have $q^{(N_A)}\geq 0$ and $\sum_{N_A}q^{(N_A)} =1$) of the (non-negative) particle number expectation values $\mbox{Tr}_{\mathcal{F}^{(HCB)}_A}\big[b_{\varphi}^\dagger b_{\varphi} \rho_A^{(N_A)}\big]$ and that all $\rho_A^{(N_A)}$ are independent in the sense that each configuration $\bd{i}_A\cup \bd{i}_B$ in Eq.~(\ref{PsiAB}) contributes to exactly one $\rho_A^{(N_A)}$. Hence, the maximum of $N^{(\varphi)}(\ket{\Psi})$ is obtained by maximizing each expectation value $\mbox{Tr}_{\mathcal{F}^{(HCB)}_A}\big[b_{\varphi}^\dagger b_{\varphi} \rho_A^{(N_A)}\big]$
separately and then picking the largest one (by choosing all other weights $q^{(N_A)}$ equal zero).
The first part of this task is already done: According to Theorem \ref{thm1}, $\mbox{Tr}_{\mathcal{F}^{(HCB)}_A}\big[b_{\varphi}^\dagger b_{\varphi} \rho_A^{(N_A)}\big]$
attains its maximum $N_{max}^{(N_A,d_A)}$ when the one-particle state $\ket{\varphi}\in \mathcal{H}_1(\mathcal{L}_A)\leq \mathcal{H}_1(\mathcal{L})$ (recall Eq.~(\ref{phiA})) is unbiased with respect to the lattice site states $\{\ket{k}\}_{k=1}^{d_A}$ and when the corresponding state
$\rho_A^{(N_A)}$ is pure, $\rho_A^{(N_A)}=\ket{\Psi}_{A\,A}\bra{\Psi}$, with $\ket{\Psi}_A$ given by Eq.~(\ref{eq:lambdaMaxState}).
Consequently, the maximal possible particle number expectation value within the lattice $\mathcal{L}_A$ is given by
\begin{equation}\label{eq:lambdaMax3}
  N_{max}^{(\mathcal{L}_A)}\equiv \max_{N_A^{-}\leq N_A \leq N_A^{+}}\big[N_{max}^{(N_A,d_A)}\big]\,,
\end{equation}
where $N_A^{\pm}$ are given by Eq.~(\ref{NAbounds}).
The total maximizer state (\ref{PsiAB}) takes the form
\begin{equation}
\ket{\Psi_{max}^{(\mathcal{L}_A)}} = \mathcal{S}_N\left[\ket{\Psi_{max}}_A\otimes \ket{N-\overline{N}_A}_B\right]\,,
\end{equation}
where $\overline{N}_A$ is the particle number maximizing (\ref{eq:lambdaMax3}), $\ket{N-\overline{N}_A}_B$ any arbitrary state of $N-\overline{N}_A$ HCB on the complementary lattice $\mathcal{L}\setminus \mathcal{L}_A$ and $\mathcal{S}_N$ denotes the symmetrizing operator for $N$ particles.

Theorem \ref{thm2}, particularly the form (\ref{PsiABmax}) of the maximizer state, shows that a locally (i.e.~within $\mathcal{L}_A$) maximal possible occupation number requires that the mode-reduced density operator $\rho_B$ of the complementary system $\mathcal{L}_B\equiv \mathcal{L}\setminus\mathcal{L}_A$ is pure, i.e.~its entanglement entropy is minimal (zero). This suggests that the entanglement entropy of the complementary system is reciprocally related to the largest occupation number within $\mathcal{L}_A$.

\bibliography{Ref_HCB}

\end{document}